\documentclass[conference]{IEEEtran}
\IEEEoverridecommandlockouts

\usepackage{cite}
\usepackage{amsmath,amssymb,amsfonts}
\usepackage{mathrsfs,amsthm}
\usepackage[ruled,linesnumbered]{algorithm2e}
\usepackage{algorithmic}
\usepackage{graphicx}
\usepackage{textcomp}
\usepackage{xcolor}
\usepackage{booktabs}
\usepackage{comment}
\usepackage{subcaption}
\def\BibTeX{{\rm B\kern-.05em{\sc i\kern-.025em b}\kern-.08em
		T\kern-.1667em\lower.7ex\hbox{E}\kern-.125emX}}

\newtheorem{Lemma}{Lemma}

\makeatletter
\newcommand{\linebreakand}{%
\end{@IEEEauthorhalign}
\hfill\mbox{}\par
\mbox{}\hfill\begin{@IEEEauthorhalign}
}
\makeatother

\begin{document}

\title{Martingale-Based Resource Partitioning Bounds for URLLC and eMBB Network Slicing in 5G\\
\thanks{This work is sponsored by Shandong Provincial Natural Science Foundation, China under Grant ZR2023QF084.}}

\author{\IEEEauthorblockN{\textit{1st} Hanbing Li}
	\IEEEauthorblockA{\textit{School of Computer Science and Artificial Intelligence} \\
		\textit{Shandong Normal University}\\
		Ji'nan, 250358, China \\
		2026317277@stu.sdnu.edu.cn}
    \linebreakand
    \IEEEauthorblockN{\textit{2nd} Wenyang Ji}
	\IEEEauthorblockA{\textit{Aerospace Information Research Institute and the Key Laboratory of Target Cognition and Application Technology (TCAT)} \\
		\textit{Chinese Academy of Sciences}\\
		Beijing, 100190, China \\
	    jiwy@aircas.ac.cn}
	\linebreakand
    \IEEEauthorblockN{\textit{3rd} Yi Zhou}
	\IEEEauthorblockA{\textit{School of Computer Science and Artificial Intelligence} \\
		\textit{Shandong Normal University}\\
		Ji'nan, 250358, China \\
		2025317128@stu.sdnu.edu.cn
}
    \and
    \IEEEauthorblockN{\textit{4th} Hao Ma}
	\IEEEauthorblockA{\textit{School of Communication and Electronic Engineering} \\
		\textit{Shandong Normal University}\\
		Ji'nan, 250358, China \\
		mahaogn@foxmail.com}
    \linebreakand
	\IEEEauthorblockN{\textit{5th} Yantong Wang$^*$}	
	\IEEEauthorblockA{\textit{School of Computer Science and Artificial Intelligence} \\
		\textit{Shandong Engineering Research Center for Multimodal Computing and Intelligent Decision Making}\\
		\textit{Shandong Normal University}\\
		Ji'nan, 250358, China \\
		yantong@sdnu.edu.cn}
}

\maketitle

\begin{abstract}
Network slicing enables the coexistence of ultra-reliable low-latency communications (URLLC) and enhanced mobile broadband (eMBB) in 5G networks. However, provisioning limited radio resources under stochastic traffic and wireless service while satisfying heterogeneous quality-of-service (QoS) requirements remains challenging. This paper proposes a martingale-based resource block (RB) allocation framework for coexisting URLLC and eMBB slices. Under the stated independence assumptions, exponential martingales are constructed using the moment generating functions of the arrival and service processes. By applying Doob's maximal inequality, we derive analytical upper bounds on the URLLC virtual-delay violation probability and the eMBB backlog violation probability. These bounds provide sufficient QoS conditions for resource allocation. A bisection procedure determines the martingale parameters, followed by an integer search for the minimum RB allocations certified by the derived bounds. Numerical results show that the proposed bounds are tighter than an optimized Chernoff--union baseline in the evaluated scenario. 
\end{abstract}


\section{Introduction}
Network slicing supports diverse service requirements in 5G and beyond networks through flexible customization and logical isolation \cite{10558834}. The coexistence of ultra-reliable low-latency communications (URLLC) and enhanced mobile broadband (eMBB) requires efficient resource allocation to meet their distinct quality-of-service (QoS) requirements \cite{10360109}. Specifically, URLLC demands low latency and high reliability, whereas eMBB primarily targets high throughput \cite{KUMAR2024208}. Allocating limited radio resources while satisfying statistical QoS requirements is therefore an important problem in network slicing \cite{LEE2024102389}. Random traffic arrivals and wireless service make queue backlogs and delays fluctuate over time. This motivates deriving bounds on their violation probabilities to guide resource block (RB) allocation between URLLC and eMBB slices.

Existing studies employ analytical methods, such as effective capacity (EC) and stochastic network calculus (SNC), as well as Monte Carlo simulations to analyze statistical QoS under stochastic traffic. EC relates wireless service statistics to queueing QoS \cite{10989629} and has been used to analyze delay violation probabilities \cite{s25041164}. Although EC provides a tractable approach for statistical QoS analysis, asymptotic approximations alone do not guarantee that delay or backlog violation probabilities remain below specified targets at a given finite time instant. Stochastic network calculus approaches based on moment generating functions (MGFs) provide analytical QoS bounds, but using union bounds to combine multiple violation events carries the risk of producing loose bounds\cite{10684888}. Such conservatism may further affect resource allocation efficiency when these bounds are used to determine resource requirements. Meanwhile, standard Monte Carlo simulations provide empirical estimates of QoS performance, but require a large number of samples to reliably estimate rare-event probabilities, making them challenging to apply under stringent reliability requirements \cite{Bai_Lam_2025}. Therefore, existing approaches face challenges in simultaneously providing finite-time QoS guarantees and efficient resource allocation under stringent reliability requirements. These limitations motivate the use of finite-time violation probability bounds to determine the minimum integer RB allocations certified by those bounds.

To address the aforementioned issues, this paper proposes a martingale-based RB allocation framework for coexisting URLLC and eMBB slices. The main contributions are summarized as follows. First, under the stated independence assumptions, we construct exponential martingales from the MGFs of the arrival and service processes and apply Doob's maximal inequality to derive non-asymptotic upper bounds on URLLC virtual-delay and eMBB backlog violation probabilities. Second, we translate these bounds into sufficient QoS conditions and develop a bisection procedure for determining the martingale parameters, followed by an integer search for the minimum RB allocations certified by the bounds. Finally, we evaluate the analytical bounds through Monte Carlo simulations and compare the resulting resource requirements with an optimized Chernoff--union baseline under identical arrival and service models. 

\section{System Model}

\subsection{Network Scenario}
\begin{figure}[htbp]
    \centering
    \includegraphics[width=0.85\linewidth]{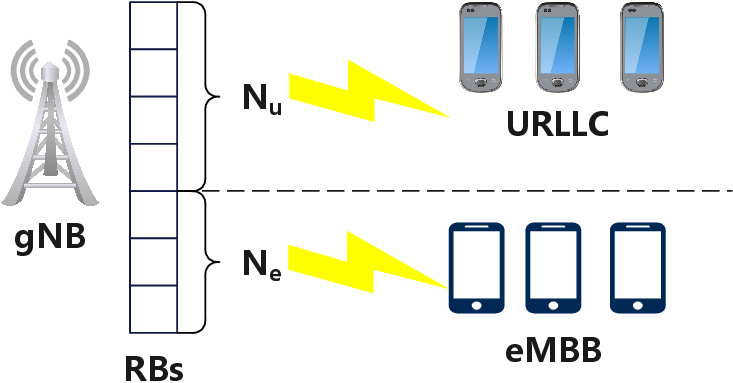}
    \caption{Network Scenario.}
    \label{fig:scenario}
\end{figure}
As shown in figure \ref{fig:scenario}, we consider the downlink of a single-cell orthogonal frequency division multiple access (OFDMA) system, in which a 5G base station (gNB) serves $K_u$ URLLC users and $K_e$ eMBB users over a total bandwidth of $B$~Hz. The bandwidth is divided into $N$ resource blocks, each of bandwidth $\Delta f = B/N$~Hz. Network slicing partitions the RBs into two slices, in which the URLLC slice is allocated $N_u$ RBs with bandwidth $B_u = N_u \Delta f$, while the eMBB slice is scheduled $N_e$ RBs with bandwidth $B_e = N_e \Delta f$, and there is $N_u + N_e \le N$. A time-slotted structure is adopted with duration $T_s$ for slice-level scheduling. Moreover, we assume block fading, i.e., the channel gain is constant within one time slot and varies independently across slots.

\subsection{Traffic Model}
Due to the heterogeneous traffic characteristics in 5G networks, URLLC traffic is modeled as discrete packet arrivals, whereas eMBB traffic is represented by its data volume in bits \cite{9732349}. For consistent queueing analysis, the URLLC packet arrivals are converted into bits using the packet size, so that the aggregate arrival processes of both slices are expressed in a common unit \cite{s21041308}.

We assume the number of packets sending to URLLC user $k \in \mathcal{U}_u$ follows a two-state ON-OFF process. In the ON state, packets arrive at a constant rate $\lambda_u$ (packets/s); in the OFF state, no packet arrives. In time slot $t$, the number of arriving packets of user $k$ is 
\begin{equation}
a_{u,k}(t) = \begin{cases} \lambda_u T_s, & \text{state ON}; \\ 0, & \text{state OFF}, \end{cases}
\label{eq:onoff}
\end{equation}
with steady-state ON probabilities $\pi_{\mathrm{on}}$ and OFF $1 - \pi_{\mathrm{on}}$, respectively. 
In addition, we assume that the ON/OFF states are independent across slots. Each URLLC packet consists of $L_u$ bits, and the aggregate URLLC arrival in slot $t$ (in bits) is 
\begin{equation}
    A_u(t) = L_u \sum_{k=1}^{K_u} a_{u,k}(t).
\end{equation}
The MGF of the per-user per-slot arrival (in bits) becomes
\begin{equation}
\mathbb{E}\left[ e^{\theta a_{u,k}(t) L_u} \right] = \pi_{\mathrm{on}}\, e^{\theta \lambda_u T_s L_u} + (1 - \pi_{\mathrm{on}}).
\end{equation}
For $K_u$ i.i.d.\ users, the aggregate MGF is 
\begin{equation}
    \varphi_{A,u}(\theta) = \left[\pi_{\mathrm{on}}\, e^{\theta \lambda_u T_s L_u} + (1 - \pi_{\mathrm{on}})\right]^{K_u}.
\end{equation}

Each eMBB user $k \in \mathcal{U}_e$ receives traffic at a mean rate $\lambda_e$ (bits/s). The amount of traffic for user $k$ in time slot $t$ follows a Poisson distribution with mean $\lambda_e T_s$, whose MGF of the per-user per-slot arrival is
\begin{equation}
\mathbb{E}\left[ e^{\theta a_{e,k}(t)} \right] = \exp\left( \lambda_e T_s \left( e^{\theta} - 1 \right) \right).
\end{equation}
For $K_e$ i.i.d.\ users, the aggregate MGF is
\begin{equation}
    \varphi_{A,e}(\theta) = \exp\left(K_e\lambda_e T_s \left( e^{\theta} - 1 \right) \right).
\end{equation}

\subsection{Service Model}
Each slice $i\in\{u,e\}$ pools the arrivals of its $K_i$ users into a single FIFO queue, for which we derive a slice-level QoS bound. Under the Shannon model, one RB provides the following service, in bits, during slot $t$:
\begin{equation}
c_i(t) = \Delta f T_s \log_2 \left( 1 + \gamma_i(t) \right),
\label{eq:shannon}
\end{equation}
where $\gamma_i(t)$ denotes the instantaneous signal-to-noise ratio (SNR) of the slice $i$ in slot $t$. Under block fading, we model $\gamma_i(t) = \rho_i g_i(t)$, where $g_i(t)$ is the normalized channel power gain with $\mathbb{E}[g_i(t)] = 1$ considering the exponential distribution for Rayleigh fading and $\rho_i$ is the average SNR.

For simplicity, we assume $N_i$ RBs for slice $i$ share the same gain $g_i(t)$,  and the total service available to the slice becomes

\begin{equation}
S_i(t) = N_i c_i(t) = N_i \Delta f T_s \log_2 \left( 1 + \rho_i g_i(t) \right).
\label{eq:service}
\end{equation}

\begin{Lemma}
\label{lem:service-mgf}
Under Rayleigh fading with normalized channel gain i.i.d. $g_i(t) \sim \mathrm{Exp}(1)$, the service MGF becomes
\begin{equation}
\varphi_{S,i}(\theta_i,N_i) = e^{1/\rho_i} \rho_i^{-\alpha_i \theta_i}\, \Gamma\!\left( 1 - \alpha_i \theta_i,\, 1/\rho_i \right),
\label{eq:mgf-service}
\end{equation}
where the upper incomplete gamma function is defined as $\Gamma(a,x)=\int_x^\infty t^{a-1}e^{-t}\,dt$ and $\alpha_i = N_i \Delta f T_s / \ln 2$.
\end{Lemma}
\begin{proof}
From eq.\eqref{eq:service}, we have $e^{-\theta_i S_i(t)} = (1 + \rho_i g_i(t))^{-\alpha_i \theta_i}$ with $\alpha_i = N_i \Delta f T_s/ \ln 2$. Since $g_i(t)$ is i.i.d., we omit the slot index $t$ hereafter. Then 
\begin{align}
    \varphi_{S,i}(\theta_i,N_i)&=\mathbb{E}\!\left[e^{-\theta_i S_i(t)}\right]
    =\mathbb{E}\!\left[(1+\rho_i g_i)^{-\alpha_i\theta_i}\right]\notag\\
    &=\int_0^\infty (1+\rho_i g)^{-\alpha_i\theta_i}e^{-g}\,dg\label{eq:lemma1}\\
    &=\frac{e^{1/\rho_i}}{\rho_i}\int_1^\infty u^{-\alpha_i\theta_i}e^{-u/\rho_i}\,du\label{eq:lemma2}\\
    &=e^{1/\rho_i}\rho_i^{-\alpha_i\theta_i}\int_{1/\rho_i}^\infty v^{(1-\alpha_i\theta_i)-1}e^{-v}\,dv.\label{eq:lemma3}
\end{align} 
Eq.\eqref{eq:lemma1} is the definition of expectation with $f_{g_i}(g)=e^{-g}$; eq.\eqref{eq:lemma2} holds considering $u=1 + \rho_i g$; and eq.\eqref{eq:lemma3} applies the substitution $v=u/\rho_i$. To align with the form $\Gamma(a,x)$, let $a\!=\!1\!-\!\alpha_i\theta_i$, $x=1/\rho_i$ yield eq.\eqref{eq:mgf-service}. Notice $1/\rho_i>0$ holds. 
\end{proof}

\subsection{QoS Requirements}
For URLLC, we consider the slot-boundary virtual delay. Let $B_u(t)$ denote the backlog after the arrivals and service in slot $t$. Under FIFO service, $W_u(t)$ is the number of additional slots required to serve all data remaining at that boundary, defined as
\[
W_u(t)=\inf\left\{d\in\mathbb{Z}_{+}: B_u(t)\leq\sum_{m=1}^{d}S_u(t+m)\right\},
\]
where the sum is zero for $d=0$. The URLLC delay violation probability should satisfy
\begin{equation}
  \Pr\left\{W_u(t)>d_u\right\}\leq\varepsilon_u,
\label{eq:qos-urllc}
\end{equation}
where $d_u$ is the delay budget in slots and $\varepsilon_u$ is the target violation probability.

Unlike latency-critical URLLC, eMBB primarily targets high throughput. We characterize its queueing QoS through a backlog constraint:
\begin{equation}
  \Pr\left\{B_e(t)> Q_e\right\}\leq\varepsilon_e,
\label{eq:qos-embb}
\end{equation}
where $B_e(t)$ is the backlog of the eMBB slice after slot $t$, $Q_e$ is the maximum permitted backlog, and $\varepsilon_e$ is the target backlog violation probability. A violation occurs only when the backlog exceeds $Q_e$.
\section{Martingale-Based QoS Bounds}

\subsection{Martingale Construction}

 {For each network slice $i \in \{u, e\}$, we assume the arrival process $A_i(t)$ and the service process $S_i(t)$ are mutually independent within each time slot $t$. Furthermore, the joint random vectors $\{(A_i(t), S_i(t))\}_{t \ge 1}$ are independent and identically distributed (i.i.d.) across time slots. All required MGFs are assumed to be finite.}




\textbf{Lemma 2.} \textit{Let $\Theta_i = \{\theta_i > 0 : \varphi_{A,i}(\theta_i)\varphi_{S,i}(\theta_i,N_i) \le 1\}$ denote the martingale domain of slice $i$. Then $\Theta_i$ is non-empty if and only if $\mathbb{E}[A_i(t)] < \mathbb{E}[S_i(t)]$.}

\begin{proof}
Let $f(\theta_i,N_i) = \ln \varphi_{A,i}(\theta_i) + \ln \varphi_{S,i}(\theta_i,N_i)$, which is strictly convex and satisfies $f(0,N_i) = 0$ and $f'(0,N_i) = \mathbb{E}[A_i(t)] - \mathbb{E}[S_i(t)]$. If $\mathbb{E}[A_i(t)] < \mathbb{E}[S_i(t)]$, then $f'(0,N_i) < 0$, so $f(\epsilon,N_i) < 0$ for sufficiently small $\epsilon > 0$, implying $\epsilon \in \Theta_i$. Conversely, if $f'(0,N_i) \ge 0$, strict convexity implies $f(\theta_i,N_i) > f'(0,N_i)\theta_i \ge 0$ for all $\theta_i > 0$, so $\varphi_{A,i}(\theta_i)\varphi_{S,i}(\theta_i,N_i) > 1$ and $\Theta_i = \emptyset$. \end{proof}

To analyze the queue backlog, consider the backlog $B_i(t)$ of slice $i$ at time slot $t$, whose evolution follows the one-step Lindley recursion:
\begin{equation}
B_i(t) = \max\left(0, \, B_i(t-1) + A_i(t) - S_i(t)\right).
\end{equation}
By recursively expanding this relation backward to the initial time slot $0$, and assuming the system starts with an empty queue (i.e., $B_i(0) = 0$), we obtain
\begin{equation}
B_i(t) = \max \left(0, \max_{1 \le k \le t} \sum_{m=t-k+1}^t (A_i(m) - S_i(m)) \right).
\end{equation}
For a given time slot $t \ge 1$, define the time-reversed cumulative net arrival process for $1 \le k \le t$ as $R_i(k) = \sum_{m=t-k+1}^t (A_i(m) - S_i(m))$ (with $R_i(0) = 0$). Since we assume an empty initial queue $B_i(0)=0$, the queue backlog can be compactly expressed as:
\begin{equation}
\label{eq:exact_backlog}
B_i(t) = \max_{0 \le k \le t} R_i(k).
\end{equation}
Furthermore, since the arrival process $A_i(t)$ and the service process $S_i(t)$ are mutually independent within each time slot, the MGF of the single-slot net arrival $A_i(t)-S_i(t)$ can be factorized as
\begin{equation*}
\mathbb{E}\left[e^{\theta_i(A_i(t)-S_i(t))}\right] = \varphi_{A,i}(\theta_i)\varphi_{S,i}(\theta_i,N_i).
\end{equation*}
Accordingly, we let $c_i = \varphi_{A,i}(\theta_i)\varphi_{S,i}(\theta_i,N_i) > 0$.

\textbf{Lemma 3.} For a given time slot $t \ge 1$ and any $\theta_i \in \Theta_i$, the process
\begin{equation}
X_i(k) = e^{\theta_i [ R_i(k)]} c_i^{-k}, \quad 0 \le k \le t 
\end{equation}
\textit{is a non-negative supermartingale with respect to the backward filtration $\tilde{\mathcal{F}}_k = \sigma(A_i(t), S_i(t), \dots, A_i(t-k+1), S_i(t-k+1))$ (where $\tilde{\mathcal{F}}_0 = \{\emptyset, \Omega\}$).}

\begin{proof}
 By definition, for $1 \le k \le t$, $R_i(k) = R_i(k-1) + A_i(t-k+1) - S_i(t-k+1)$. Since  $R_i(k-1)$ is $\tilde{\mathcal{F}}_{k-1}$-measurable, and $(A_i(t-k+1), S_i(t-k+1))$ is independent of $\tilde{\mathcal{F}}_{k-1}$ due to the independent and identically distributed (i.i.d.) nature across time slots, we have
\begin{equation*}
\begin{aligned}
&\mathbb{E}[X_i(k) \mid \tilde{\mathcal{F}}_{k-1}] 
= c_i^{-k} e^{\theta_i [R_i(k-1)]} \cdot \mathbb{E}\left[e^{\theta_i A_i(t-k+1)}\right] \\
&\quad \cdot\mathbb{E}\left[e^{-\theta_i S_i(t-k+1)}\right] = c_i^{-k} e^{\theta_i [R_i(k-1)]} c_i = X_i(k-1).
\end{aligned}
\end{equation*}
Since $X_i(k) > 0$ and its expectation is finite (i.e., integrability holds with 
$\mathbb{E}[X_i(k)] = \mathbb{E}[X_i(0)] = 1 < \infty$), thus it satisfies the martingale 
equality. Therefore it can be viewed as a non-negative supermartingale 
with initial state $X_i(0) = 1$.
\end{proof}

\subsection{Backlog and Delay Violation Bounds}

\textbf{Theorem 1.} \textit{For slice $i \in \{u, e\}$ and any $\theta_i \in \Theta_i$, the queue backlog violation probability bound is given by}

\begin{equation}
\label{eq:theorem1_bound}
P(B_i(t)> Q_i) \le e^{-\theta_i Q_i}, \quad \forall Q_i \ge 0.
\end{equation}

\begin{proof}
From eq.\eqref{eq:exact_backlog} we have $B_i(t) = \max_{0 \le k \le t} R_i(k)$. Since $c_i \le 1$ for any $\theta_i \in \Theta_i$ (which implies $c_i^{-k} \ge 1$ for all $0 \le k \le t$), we obtain the following event inclusions:
\begin{equation*}
\begin{aligned}
&\{B_i(t) > Q_i\} = \left\{\max_{0 \le k \le t} R_i(k) > Q_i\right\} \\
&= \left\{\max_{0 \le k \le t} e^{\theta_i[ R_i(k)]} > e^{\theta_i Q_i}\right\}\subseteq \left\{\max_{0 \le k \le t} X_i(k) \ge e^{\theta_i Q_i}\right\}.
\end{aligned}
\end{equation*}
Applying Doob's maximal inequality to the non-negative supermartingale $X_i(k)$ yields
\begin{equation*}
\begin{aligned}
P(B_i(t) > Q_i) &\le P\left(\max_{0 \le k \le t} X_i(k) \ge e^{\theta_i Q_i}\right) \\
&\le \mathbb{E}[X_i(0)] e^{-\theta_i Q_i} = e^{-\theta_i Q_i},
\end{aligned}
\end{equation*}
which establishes the theorem. 
\end{proof}


\textbf{Theorem 2.} \textit{For any $\theta_u \in \Theta_u$, the virtual-delay violation probability bound of the URLLC slice is given by}
\begin{equation}
\label{eq:theorem2_bound}
P(W_u(t) > d_u) \le [\varphi_{S,u}(\theta_u, N_u)]^{d_u}.
\end{equation}

\begin{proof}
 By the definition of the slot-boundary virtual delay, $\{W_u(t)>d_u\}=\{B_u(t)>Z\}$, where $Z=\sum_{m=t+1}^{t+d_u}S_u(m)$ is the cumulative service over the subsequent $d_u$ slots. Under the assumed independence across slots, $B_u(t)$ depends on the history up to slot $t$ while $Z$ depends on future slots, so $B_u(t)$ is independent of $Z$. Conditioning on $Z=z$ with probability density function $f_Z(z)$ and applying Theorem 1, we obtain
\begin{equation*}
\begin{aligned}
P(W_u(t)>d_u)
&= \int_0^\infty P(B_u(t)>z)f_Z(z)\,dz \\
&\le \int_0^\infty e^{-\theta_u z}f_Z(z)\,dz = \mathbb{E}[e^{-\theta_u Z}],
\end{aligned}
\end{equation*}
where the inequality follows from Theorem~1, which gives $P(B_u(t)>z)\le e^{-\theta_u z}$. Since the service process $\{S_u(m)\}$ is i.i.d.\ across time slots, the transform of the sum factorizes into the product of individual service transforms, yielding $\mathbb{E}[e^{-\theta_u Z}]=[\varphi_{S,u}(\theta_u,N_u)]^{d_u}$.
\end{proof}




\subsection{Bandwidth Allocation}

Combining the violation probability bounds in Theorems 1 and 2 with the slice QoS targets in  eq.\eqref{eq:qos-urllc}  and eq.\eqref{eq:qos-embb}, we obtain the following resource allocation result.

\textbf{Theorem 3.} \textit{Let $\Theta_i(N_i)$ denote the martingale domain for a given allocation $N_i$. The minimum numbers of RBs that guarantee the URLLC and eMBB QoS are}
\begin{align}
N_u^* &= \min \!\left\{ N_u \!: \!\max_{\theta_u \in \Theta_u(N_u)}\! \big[\!-\!d_u \ln \varphi_{S,u}(\theta_u, N_u)\big] \!\ge \ln \frac{1}{\epsilon_u} \right\}, \label{eq:Nu_star} \\
N_e^* &= \min\! \left\{ N_e \!:\! \max_{\theta_e \in \Theta_e(N_e)} \!\theta_e Q_e \ge \ln \frac{1}{\epsilon_e} \right\},\label{eq:Ne_star}
\end{align}
\textit{and the corresponding minimum bandwidth demands are $B_u^* = N_u^* \Delta f$ and $B_e^* = N_e^* \Delta f$.}

\begin{proof}
From eq.\eqref{eq:theorem2_bound}, the URLLC delay requirement eq.\eqref{eq:qos-urllc} is satisfied whenever there exists $\theta_u \in \Theta_u(N_u)$ such that $[\varphi_{S,u}(\theta_u, N_u)]^{d_u} \le \epsilon_u$. Taking the natural logarithm on both sides and rearranging the terms yields $-d_u \ln \varphi_{S,u}(\theta_u, N_u) \ge \ln(1/\epsilon_u)$, which leads to eq.\eqref{eq:Nu_star}. The eMBB allocation in eq.\eqref{eq:Ne_star} follows analogously from eq.\eqref{eq:theorem1_bound} and eq.\eqref{eq:qos-embb}.
\end{proof}


\section{Resource Partitioning Algorithm}

Theorem 3 decouples the statistical QoS requirements of the two slices into independent minimum RB requirements $N_u^*$ and $N_e^*$. Since both objective functions inside the maxima in eq.\eqref{eq:Nu_star} and eq.\eqref{eq:Ne_star} are strictly increasing with respect to $\theta_i > 0$, their greatest quantities are inevitably attained at the boundary of the martingale domain, i.e., $\theta_i^*(N_i)\!=\!\max\{\theta_i\! >\! 0: \!\varphi_{A,i}(\theta_i)\varphi_{S,i}(\theta_i, N_i)\!\le\!1\}$. The resource partitioning problem then evaluates whether the available $N$ RBs can accommodate both slices based on $\theta_i^*(N_i)$. If no allocation satisfies these sufficient conditions within the available capacity, the derived bounds cannot certify both QoS targets.

As proved in {Lemma 2}, $f(\theta_i,N_i)$ is strict convex in $\theta_i$ with $f(0,N_i)=0$ and $f'(0,N_i)<0$ in the stable regime; it first decreases below $0$ and then rises back to $0$ at $\theta_i^*(N_i)$ by verifying $\lim_{\theta_i\rightarrow\infty}f(\theta_i,N_i)=\infty$, making bisection applicable. The bisection procedure is summarized in Algorithm~\ref{alg:bisection}.
\begin{algorithm}[htbp]
\caption{Bisection for $\theta_i^*(N_i)$}
\label{alg:bisection}
\begin{algorithmic}[1]
\REQUIRE Allocation $N_i$, tolerance $\delta$, start-point $\theta_0$
\ENSURE Optimal parameter $\theta_i^*(N_i)$ given $N_i$
\STATE $\theta_{\mathrm{low}}\gets 0$, $\theta_{\mathrm{high}}\gets \theta_0$
\WHILE{$f(\theta_{\mathrm{high}};N_i)\le 0$}
    \STATE $\theta_{\mathrm{high}}\gets 2\,\theta_{\mathrm{high}}$ 
\ENDWHILE
\WHILE{$\theta_{\mathrm{high}}-\theta_{\mathrm{low}}>\delta$}
    \STATE $\theta_{\mathrm{mid}}\gets(\theta_{\mathrm{low}}+\theta_{\mathrm{high}})/2$
    \IF{$f(\theta_{\mathrm{mid}};N_i)\le 0$}
        \STATE $\theta_{\mathrm{low}}\gets\theta_{\mathrm{mid}}$
    \ELSE
        \STATE $\theta_{\mathrm{high}}\gets\theta_{\mathrm{mid}}$
    \ENDIF
\ENDWHILE
\RETURN $(\theta_{\mathrm{low}}+\theta_{\mathrm{high}})/2$
\end{algorithmic}
\end{algorithm}

Since increasing $N_i$ reduces the service MGF and hence enlarges $\Theta_i(N_i)$, the sets in {eq.(21)--eq.(22)} are monotone and the minima are well defined. Then the minimum $N_i^*$ can be found by incrementally increasing $N_i$ and checking the QoS constraint. Define $F_u(N_u)=-d_u\ln\varphi_{S,u}(\theta_u^*,N_u)$ and $F_e(N_e)=\theta_e^*Q_e$. The sufficient QoS condition for slice $i$ is then $F_i(N_i)\ge\ln(1/\varepsilon_i)$. The allocation algorithm is given in Algorithm~\ref{alg:rb_allocation}.

\begin{algorithm}[htbp]
\caption{RB Allocation for URLLC \& eMBB}
\label{alg:rb_allocation}
\begin{algorithmic}[1]
\REQUIRE QoS targets $(d_u,\varepsilon_u)$, $(Q_e,\varepsilon_e)$; SNR $\rho_u,\rho_e$; total capacity $N$
\ENSURE Minimum bound-certified allocations $N_u^*,N_e^*$, or \textsc{Not Certified}
\FOR{each slice $i\in\{u,e\}$}
    \STATE Calculate $N_{i,\min}=\lfloor \mathbb{E}[A_i]/\mathbb{E}[c_i(t)]\rfloor+1$
    \STATE $N_i\gets N_{i,\min}$, $\mathrm{found}\gets\mathrm{false}$
    \WHILE{$N_i\le N$}
        \STATE Obtain the martingale parameter via Algorithm~\ref{alg:bisection}
        \STATE Calculate $F_i(N_i)$ based on eq.\eqref{eq:Nu_star} or eq.\eqref{eq:Ne_star}
        \IF{$F_i(N_i)\ge\ln(1/\varepsilon_i)$}
            \STATE $N_i^*\gets N_i$
            \STATE $\mathrm{found}\gets\mathrm{true}$
            \STATE \textbf{break}
        \ENDIF
        \STATE $N_i\gets N_i+1$
    \ENDWHILE
    \IF{$\mathrm{found}=\mathrm{false}$}
        \RETURN \textsc{Not Certified}
    \ENDIF
\ENDFOR
\IF{$N_u^*+N_e^*\le N$}
    \RETURN $(N_u^*,N_e^*)$
\ELSE
    \RETURN \textsc{Not Certified}
\ENDIF
\end{algorithmic}
\end{algorithm}

The outer loop of Algorithm~\ref{alg:rb_allocation} iterates at most $N$ times per slice, and each bisection run requires $O(\log(1/\delta))$ evaluations of $f(\theta_i;N_i)$. Each evaluation involves one upper incomplete gamma function (Lemma~\ref{lem:service-mgf}) and one exponential (Poisson or ON--OFF arrival MGF), both $O(1)$. The total complexity per slice is $O(N\log(1/\delta))$. In practice, a small safety margin $\theta_i\leftarrow(1-\eta)\theta_i^*$ with $\eta\ll1$ is applied to avoid numerical boundary violations where $\varphi_{A,i}\varphi_{S,i}$ slightly exceeds 1 due to floating-point errors.

\section{NUMERICAL INVESTIGATIONS}
\label{sec:numerical_investigations}

\subsection{Simulation Setup}

Simulation parameters are summarized in Table~\ref{tab:simulation_parameters}. All methods use the same arrival and service models. The Chernoff--union baseline used here \cite{6868978} includes the geometric factor $1/(1-q_i)$, where $q_i=\varphi_{A,i}(\theta)\varphi_{S,i}(\theta,N_i)$. The resulting bounds are $[\varphi_{S,u}(\theta,N_u)]^{d_u}/(1-q_u)$ for URLLC virtual delay and $e^{-\theta Q_e}/(1-q_e)$ for eMBB backlog. For each RB allocation, we minimize each baseline bound over $\theta>0$ with $q_i<1$ and cap the result at one.
\begin{table}[htbp]
    \caption{Simulation Parameters\cite{9585085}}
    \label{tab:simulation_parameters}
    \centering
    \footnotesize
    \renewcommand{\arraystretch}{1.12}
    \setlength{\tabcolsep}{4pt}
    \begin{tabular}{ll}
        \hline
        \textbf{Parameter} & \textbf{Value} \\
        \hline
        System bandwidth, $B$ & $10$ MHz \\
        Total number of RBs, $N$ & $50$ \\
        Bandwidth per RB, $\Delta f$ & $200$ kHz \\
        Slot duration, $T_s$ & $0.5$ ms \\
        Average SNR, $\rho_u,\rho_e$ & $10$ dB \\
        \hline
        URLLC users, $K_u$ & $10$ \\
        URLLC ON probability, $\pi_{\mathrm{on}}$ & $0.2$ \\
        URLLC ON-state packet rate, $\lambda_u$ & $10^4$ packets/s \\
        URLLC packet size, $L_u$ & $160$ bits \\
        URLLC delay threshold, $d_u$ & $4$ slots ($2$ ms) \\
        \hline
        eMBB users, $K_e$ & $10$ \\
        eMBB arrival rate per user, $\lambda_e$ & $600$ kbit/s \\
        eMBB backlog threshold, $Q_e$ & $12$ kbits \\
        \hline
        Common target, $\varepsilon_u=\varepsilon_e$ & $10^{-1}\sim10^{-6}$ \\
        Independent runs per slice & $10$ \\
        Measured slots per run & $5\times10^7$ \\
        Warm-up slots per run & $2\times10^4$ \\
        \hline
    \end{tabular}
\end{table}

\subsection{URLLC Delay Violation Probability}
\begin{figure}[htbp]
    \centering
    \includegraphics[width=\columnwidth]{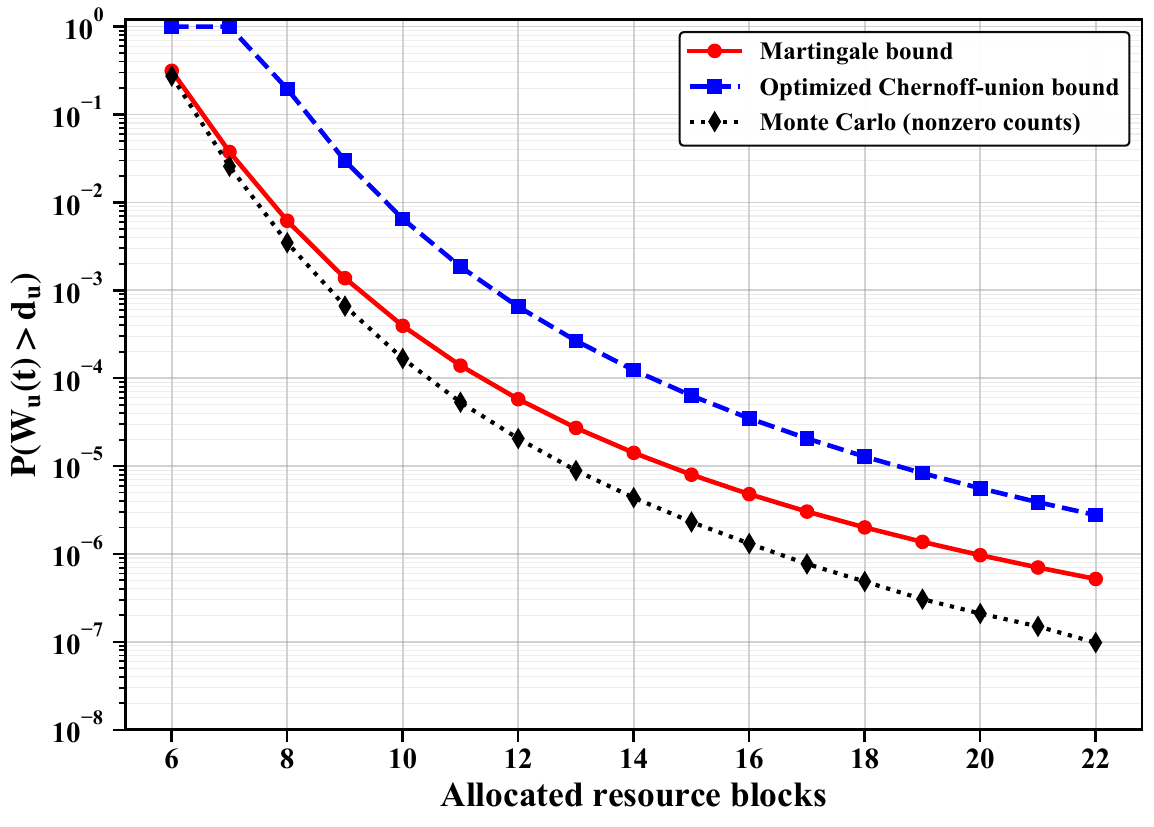}
    \caption{URLLC virtual-delay violation probability versus the allocated RBs for $d_u=4$ slots.}
    \label{fig:urllc_bounds}
\end{figure}
Figure~\ref{fig:urllc_bounds} shows the virtual-delay violation probability versus the allocated RBs. Increasing $N_u$ reduces the violation probability by providing more service within the delay budget. The martingale bound is tighter than the Chernoff--union baseline throughout the plotted range and follows the Monte Carlo estimates more closely. At $N_u=20$, the martingale bound falls below $10^{-6}$, whereas the baseline remains above this target.

\subsection{eMBB Backlog Violation Probability}
\begin{figure}[htbp]
    \centering
    \includegraphics[width=\columnwidth]{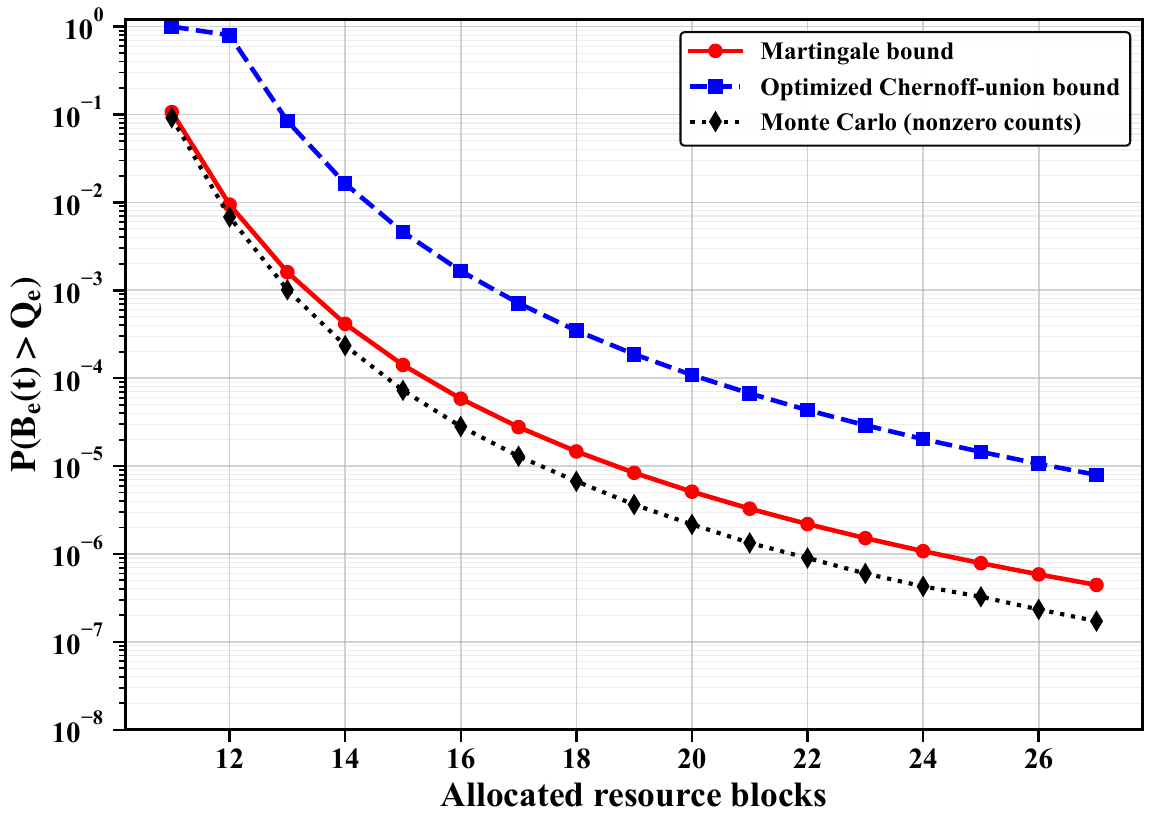}
    \caption{eMBB backlog violation probability versus the allocated RBs for $Q_e=12$~kbits.}
    \label{fig:embb_bounds}
\end{figure}
Figure~\ref{fig:embb_bounds} shows the backlog violation probability $P\left(B_e(t)>Q_e\right)$ versus the allocated RBs. Increasing $N_e$ provides more service to the eMBB queue and reduces the probability that the backlog exceeds the threshold. The martingale bound is lower than the Chernoff--union baseline throughout the plotted range and follows the Monte Carlo estimates more closely. At $N_e=24$, the martingale bound is approximately $1.08\times10^{-6}$, slightly above the target of $10^{-6}$, whereas at $N_e=25$, it falls below this target. In contrast, the baseline bound remains above the target at this allocation and therefore requires more RBs to certify the same QoS requirement. 
\subsection{Resource Allocation Comparison}
\begin{figure}[htbp]
    \centering
    \includegraphics[width=\columnwidth]{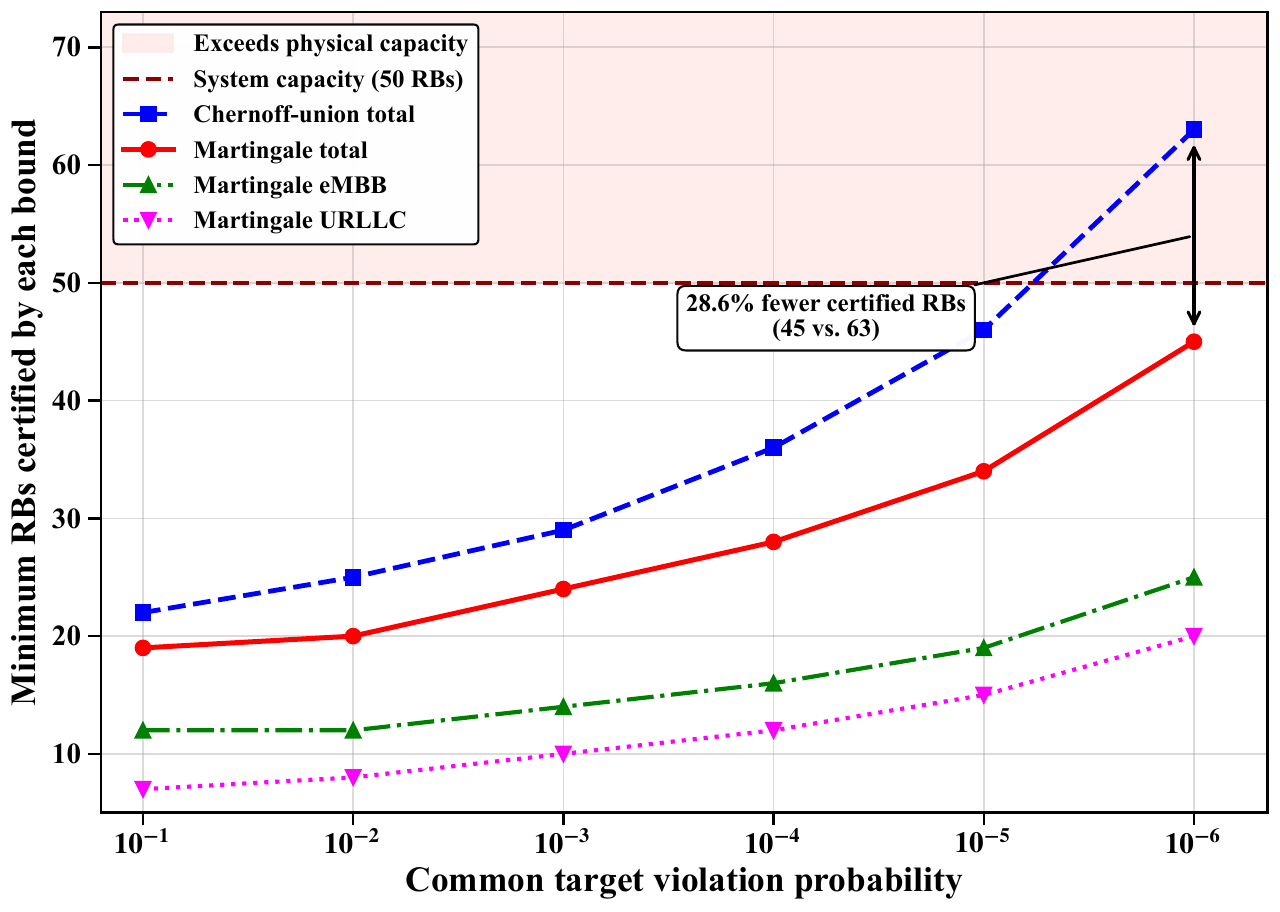}
    \caption{Minimum RB requirements certified by the martingale and optimized Chernoff--union bounds.}
    \label{fig:rb_allocation}
\end{figure}

Figure~\ref{fig:rb_allocation} compares the minimum RB requirements certified by the two bounds for a common target $\varepsilon_u=\varepsilon_e=\varepsilon$. Stricter targets increase the resource requirements of both methods. At $\varepsilon=10^{-6}$, it allocates 20 RBs to URLLC and 25 RBs to eMBB, totaling 45 RBs, whereas the baseline requires 63 RBs. Under the stated independence assumptions and the parameters in Table~\ref{tab:simulation_parameters}, this reduces the minimum certified total by $28.6\%$.
The proposed allocation remains within the 50-RB capacity, whereas the baseline cannot certify both QoS targets within this capacity.


\section{CONCLUSIONS}
\label{sec:conclusions}

This paper developed a martingale-based RB allocation framework for coexisting URLLC and eMBB slices. Under the stated independence assumptions, exponential martingale techniques yield non-asymptotic upper bounds on URLLC virtual-delay and eMBB backlog violation probabilities. These bounds enable the determination of the minimum integer RB allocations satisfying the corresponding sufficient QoS conditions.
Numerical results show that the proposed bounds are tighter than the optimized Chernoff--union baseline in the evaluated scenario. At a common violation probability target of $10^{-6}$, the proposed allocation requires 45 RBs, compared with 63 RBs for the baseline, reducing the bound-certified resource requirement by $28.6\%$ while remaining within the 50-RB system capacity. Future work will extend the framework to temporally correlated traffic and fading channels. We will also investigate dynamic RB allocation under time-varying scenarios. Furthermore, some rare-event simulation methods,  such as importance sampling and splitting, will be explored to improve the accuracy of probability estimates under more stringent reliability requirements.
\bibliographystyle{ieeetr}
\bibliography{reference}

\end{document}